\documentclass[a4paper,10pt,twoside]{cpc-hepnp}

\usepackage{multicol}
\usepackage{graphicx}
\usepackage{booktabs}
\usepackage{amssymb,bm,mathrsfs,bbm,amscd}
\usepackage[tbtags]{amsmath}
\usepackage{lastpage}
\usepackage{CJK}

\begin{document}

\fancyhead[co]{\footnotesize YANG Zhen-wei~ et al:
       Simulation Study on neutrino nucleus cross section measurement in Segmented Detector}

\footnotetext[0]{Received 14 March 2009}

\title{Simulation Study on neutrino nucleus cross section measurement in Segmented Detector at Spallation Neutron Source%
       \thanks{Supported by National Natural Science Foundation of China (10875062) }
      }

\author{%
      YANG Zhen-wei$^{1,2,3;1)}$\email{yangzhw@tsinghua.edu.cn}%
\quad DING Ming-ming$^{1,2,3}$%
\quad CHEN Shao-min$^{1,2,3}$\\
\quad WANG Zhe$^{3}$%
\quad ZHANG Feng$^{1,2,3}$%
\quad GAO Yuan-ning$^{1,2,3}$
}
\maketitle

\address{%
  $^1$ Key Laboratory of Particle \& Radiation Imaging (Tsinghua University), Ministry of Education\\
  $^2$ Department of Engineering Physics, Tsinghua University, Beijing 100084, P. R. China\\
  $^3$ Center for High Energy Physics, Tsinghua University, Beijing 100084, P. R. China\\
}

\begin{abstract}
Knowledge of $\nu_e$-$\mathrm{Fe}$/$\mathrm{Pb}$ differential cross sections for $\nu_e$ energy
below several tens of MeV scale is believed to be crucial in understanding Supernova physics.
In a segmented detector at Spallation Neutrino Source, $\nu_e$ energy reconstructed from the
electron range measurement is strongly affected because of both multiple scattering and
electromagnetic showers occurring along the electron passage in target materials.
In order to estimate the effect, a simulation study has been performed with a cube block
model assuming a perfect tracking precision. The distortion of energy spectrum is
observed to be proportional to the atomic number of target material. Feasibility of
unfolding the distorted $\nu_e$ energy spectrum is studied for both Fe and Pb cases.
Evaluation of statistical accuracy attainable is therefore provided for a segmented detector.
\end{abstract}



\begin{multicols}{2}
\section{Introduction}
\label{introduction}
Precise knowledge of the differential cross sections for quasi-elastic neutrino-nucleus scattering
at a scale up to several tens of MeV is of importance in understanding the explosion dynamics of supernova,
since the neutrinos play a decisive role
in the development between the prompt shock and the delayed shock~\cite{sn}.
Among the cross sections
neutrino interactions with $^2\mathrm{H}$, $\mathrm{C}$,
$\mathrm{O}$, $\mathrm{Fe}$ and $\mathrm{Pb}$ are most relevant.
In this low energy scale, the cross section varies rapidly,
leading to large uncertainty in predicting the core collapse supernova~\cite{sn_nu,Kuzmin_K}.
Although some theoretical calculations~\cite{xsec_th} of neutrino-nucleus cross sections
have been performed in recent years, large uncertainties still exist and need to be clarified from experiments.
On the experiment side, the differential cross sections with $\mathrm{C}$ and $\mathrm{O}$
have been measured by KARMEN and LSND experiments~\cite{xsec_karmen,xsec_lsnd},
but those with $\mathrm{Fe}$ and $\mathrm{Pb}$ are still not available.
In this paper, we present a simulation study on cross section measurement
of neutrino interaction with Fe and Pb.
The motivation is to give an estimate of the achievable limit
in the experiment of neutrino cross section measurement with a segmented detector
at Spallation Neutron Source(SNS).

At SNS, protons are driven to hit a target,
producing a large amount of neutrons and by-product pions.
The negative pions are mostly captured in the target, while the positive pions,
after having been stopped in the target within $0.1$ ns or so,
undergo the characteristic successive decay scheme
$\pi^+\xrightarrow{26\ \mathrm{ns}}\mu^+ +\nu_\mu$
and followed by $\mu^+\xrightarrow{2200\ \mathrm{ns}}e^++\nu_e+\bar{\nu}_\mu$.
The resultant $\nu_e$'s can have a flux up to $2\times10^7$ $/\mathrm{cm}^2/\mathrm{s}$ at a distance of
20 meters away from the spallation target for the SNS at the Oak Ridge National Laboratory,
assuming the SNS power to be 1 MW~\cite{Proposal_uh}.
Since the time structure of SNS beam provides a unique advantage for neutrino
studies\cite{Proposal_uh,Marschuw_R,Armbruster_B},
backgrounds are strongly suppressed.
In the SNS target, the pions are stopped and decay to muons.
The subsequent muon decay produces equal intensity of $\nu_e$ and $\bar \nu_\mu$,
with a maximum energy 52.8 MeV. The energy spectra of the three kinds of neutrinos
are shown in Fig.\ref{fig:SNSneutrino}.

Such an energy range overlaps extremely well with that of supernova neutrinos.
This characteristics makes it valuable to measure cross sections of interactions
between neutrinos and various target materials at SNS.
Fig.~\ref{fig:detector} shows the sketch of neutrino detector at SNS.
A $\nu-$SNS project\cite{Proposal_uh} was proposed to build a neutrino facility
at the SNS at Oak Ridge National Laboratory.
However, there is no report on how precise the measurement can be achieved.
\begin{center}
\includegraphics[scale=0.4]{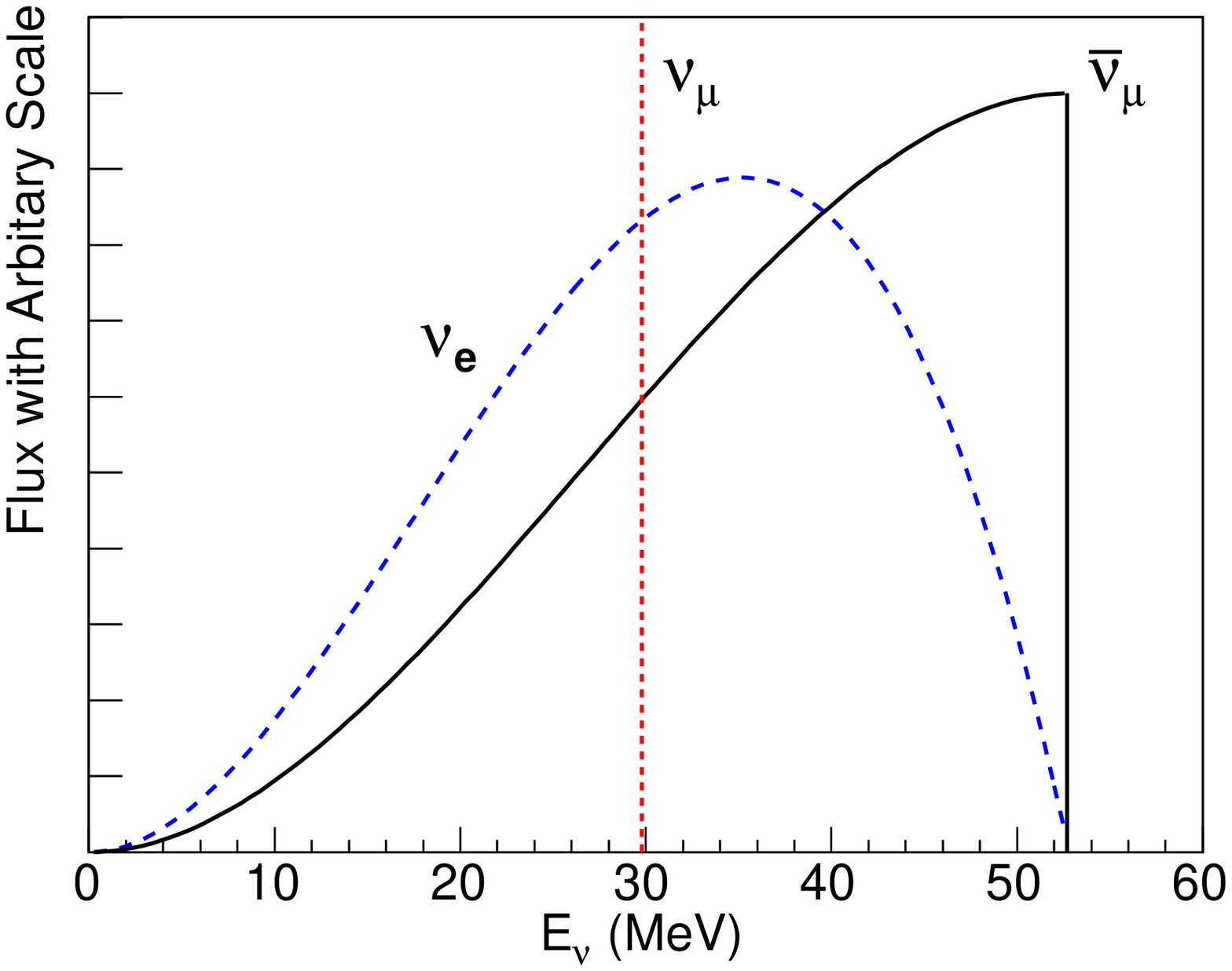}
 \figcaption{\label{fig:SNSneutrino} Neutrino spectrum from muon and pion decay-at-rest at a spallation neutron source. }
\end{center}

\begin{center}
\includegraphics[width=8 cm]{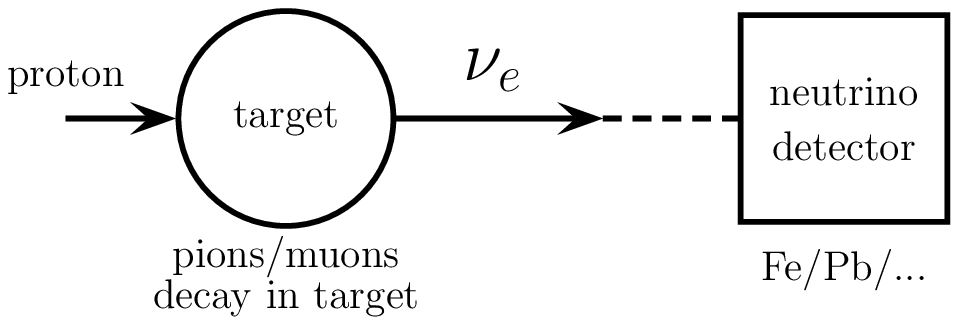}
 \figcaption{ \label{fig:detector} Sketch of neutrino detector at SNS. }
\end{center}

Electron neutrinos interacting with target material at the SNS
undergo both neutral- and charged-current interactions,
while muon neutrinos can only undergo neutral-current interactions
due to the limitation of kinematics.
The neutral-current events are suppressed by at least two orders of magnitude
in comparison with the charged-current events and can only be measured by the decay of excited nucleus.
The charged-current events can be detected via the outgoing electron
in the neutrino-nucleus reaction
$\nu_e+\ _{Z}^{A}\mathrm{X}_{N}\rightarrow\ _{Z+1}^{A}\mathrm{Y}_{N-1}+e^-$
($\nu_e+n\rightarrow p+e^-$).
Since the neutrino energy is much lower than the nucleus,
considering the charged-current quasi-elastic(CCQE) process and ignoring fermi motion of nucleus,
the neutrino energy $E_{\nu_e}$ can be determined by the induced electron energy $E_e$,
\begin{equation}
 \label{eq:energy_relation}
 E_{\nu_e} = E_e+\Delta,
\end{equation}
where $\Delta$ is the mass difference between the final state nucleus $\mathrm{Y}$
and the initial state nucleus $\mathrm{X}$.
Therefore, a precise energy reconstruction of the $\nu_e$ induced electrons
is essential to determine the differential cross sections.
The segmented detector is proposed to measure the trajectories of $\nu_e$ induced electrons
in the target.
However, the energy lose of the electron is dominated by ionization at low energy,
while above the critical energy $E_c$, it will be dominated by bremsstrahlung.
The electromagnetic shower caused by bremsstrahlung will lead to large uncertainty on
the measurement of electron track length for energy above $E_c$.
The critical energy can be approximated~\cite{Ec,PDG}
by $E_c=800\ \mathrm{MeV}/(Z+1.2)$, where $Z$ is the atomic number of the target nucleus.
Therefore the critical energy of $\mathrm{Fe}$ is $E_c=29.4\ \mathrm{MeV}$,
and $E_c=9.6\ \mathrm{MeV}$ for $\mathrm{Pb}$.
This makes it impossible to obtain the electron energy by its track length event by event.
In this paper, we will show a statistical application to unfold the distorted
energy spectrum, and leading to a reasonable resolution of the differential cross section.

 The outline of this paper is as followed.
 In section 2, the Geant4 simulation of neutrino induced electrons are simulated
 in $\mathrm{Fe}$ and $\mathrm{Pb}$.
 The simulation results and analysis are described in section 3,
 and a brief conclusion is made in section 4.

\section{Simulation procedure}
To simplify the issue, we use a cube block target material with a fiducial
volume of $1\times1\times1$ $\mathrm{m}^3$, in stead of the segmented detector with
target, tube, wire, etc. The tracking resolution is assumed to be perfect.
Geant4 (version 9.0p01) package~\cite{Geant4} is applied in this simulation study.

The cross section of neutrino-nucleus scattering is rather small,
we therefore directly simulate the behaviors of the induced electron in the target.
The input energy spectrum of electron is the Michel distribution shaped by
the theoretically predicted cross section given by R. Lazauskas and C. Volpe~\cite{xsec_th}.
The maximum energy of induced electron, $E_{\max}$,
is approximately $48.2$ MeV on $^{56}\mathrm{Fe}$,
and the threshold energy of the interaction is $E_{\mathrm{thres}}\simeq5.1$ MeV.
For $^{208}\mathrm{Pb}$, $E_{\max}\simeq49.9$ MeV, and $E_{\mathrm{thres}}\simeq3.4$ MeV.

Daughter products e.g., $^{56}\mathrm{Co}$, $^{208}\mathrm{Bi}$ are ignored,
since the life times of the daughters are long
comparing the search window ($\sim10$ micro seconds).

The material of the detector is selected to be iron or lead with natural abundances,
but not pure $^{56}\mathrm{Fe}$ or $^{208}\mathrm{Pb}$.
Therefore, the measured cross section is an average of mixture isotopes of iron or lead.
This affects iron little since the natural abundance of $^{56}\mathrm{Fe}$ is $91.754\%$.
For lead, the natural abundances of $^{208}\mathrm{Pb}$, $^{207}\mathrm{Pb}$
and $^{206}\mathrm{Pb}$ are $52.4\%$, $22.1\%$ and $24.1\%$ respectively,
the effect of the mixture should be considered seriously in later studies.

For the interaction of electrons in nuclear material,
electromagnetic process is the most important one.
The processes of multiple scattering, low energy ionization
and low energy bremsstrahlung are added for electrons in the simulation.

To perform the unfolding procedure~\cite{Cowan_G,Albert_J,Anykeev_V},
two kinds of spectra of electrons are generated:
1) Michel distribution used to obtain the detector response matrix;
2) Michel distribution weighted with the cross section of corresponding $\nu_e$-nucleus interaction.
The second one is considered as the true distribution to be reconstructed.
For each event, the electron's energy (input) $E$ and track length $L$ in the target are recorded.

The number of events attainable
in one nominal year (assuming ~$10^7$ seconds) is estimated
by considering the flux of electron neutrino, the size of detector and the average
cross section of $\nu_e$-nucleus scattering.
The expected maximum cross section at $52.8\ \mathrm{MeV}$ for $\mathrm{Fe}$ is
$\sigma_{\mathrm{Fe},\max}\simeq{}1.2\times10^{-39}\ \mathrm{cm}^2$, and it gives that
the average cross section over energy
is about $0.3\sigma_{\mathrm{Fe},\max}$.
For $\mathrm{Pb}$, $\sigma_{\mathrm{Fe},\max}\simeq{}1.3\times10^{-38}\ \mathrm{cm}^2$,
and the average cross section over energy is about $0.4\sigma_{\mathrm{Pb},\max}$.
Assuming the effective volume of target is $1\ \mathrm{m}^3$,
and the flux of $\nu_e$'s is $2\times10^7\ \mathrm{cm}^{-2}\mathrm{s}^{-1}$,
the number of events in one nominal year is estimated to be $6.1\times10^3$ for Fe target,
and $3.4\times10^4$ for Pb target.

\section{Unfolding method}
In the energy measurement, due to detector resolution, efficiency and sometimes intrinsic physical reason,
the true value in bin $i$, will usually migrate into measured energy bin $j$.
In other words, the measured distribution is usually distorted by the detector response,
or the detector response is folded to the true distribution, leading to the measured distribution.
In many cases, the true distribution could be restored by applying correction factors
to the measured distribution.
This method fails, however, if the relation between true value and measured ones
is highly nonlinear where the behavior of detector response is poor.

Suppose the true distribution is $X_{\mathrm{true}}$,
and the measured one is $Y_{\mathrm{obs}}$.
If effects of efficiency and background are ignored, they satisfy
\begin{equation}
 \label{eq:unfolding_1}
  Y_{\mathrm{obs}}=R\cdot{}X_{\mathrm{true}},
\end{equation}
where $R$ is the response matrix or the migration matrix which describes the effects of detector response.
The basic idea of unfolding is that the effects of detector response could be removed or unfolded
by inverse of the response matrix, i.e., the estimator of the true distribution $X$ is
\begin{equation}
 \label{eq:unfolding_2}
  \hat{X}_{\mathrm{true}}=R^{-1}\cdot{}Y_{\mathrm{obs}}.
\end{equation}
However, although this is the unbias estimator with smallest variance,
the physics is totally swept out by unphysical oscillation.
The regularization is needed to remove the unphysical oscillation,
which is actually a compromise between bias and unphysical oscillation,
controlled by the choice of regularization parameter.
There are several different schemes of regularisation~\cite{Cowan_G,Albert_J,Anykeev_V},
for example, Tiknonov reularizaton, regularization based on entropy,
Bayesian regularization, regularization based on Singular Value Decomposition(SVD), etc.
The basic point of all is that the true distribution
is believed to be smooth but not oscillating rapidly.
In this study, we used the TSVDUnfold class in ROOT package\cite{ROOT}.
TSVDUnfold is a class in ROOT for unfolding based on SVD technique,
it is part of RooUnfold, a ROOT unfolding framework,
developed by Tim Adye, {\it et al}~\cite{RooUnfold}.

In the SNS neutrino experiment, to measure the differential cross section
of neutrino-nucleus scattering,
the true value is the energy of incident electron, which is principally unknown
and supposed to be binned in histogram $E$,
the measured value is the electron track length in the target, binned in histogram $L$.
The response matrix $R$ can be obtained by Monte Carlo simulation
as long as we have full knowledge on the detector.
In discrete expression,
\begin{equation}
 \label{eq:unfolding_3}
 L_j = \sum_{i=1}^{n_a} R_{ji}\cdot E_i \hspace{1cm}(j=1,\ldots,n_b),
\end{equation}
where $E_i$ is number of events in bin $i$ of $E$, $L_j$ is number
of events in bin $j$, $n_a$ and $n_b$ are numbers of bins of $E$ and
$L$ respectively. The physical meaning of $R_{ji}$ is the probability of an event
in bin $i$ of true distribution to migrate into bin $j$ of measured distribution.
Therefore, the response matrix is actually the normalized two-dimensional scattering plot
of $L$ and $E$.

\section{Simulation results and analysis}
In the simulation, the electron track length $L$ is directly read
from Geant4 with no uncertainty.
The response matrix of detector with the iron target is shown in Fig.~\ref{fig:responseM}(a),
and that of detector with the lead target is shown in Fig.~\ref{fig:responseM}(b),
corresponding to 200,000 electrons with Michel distributed energy.
If the matrix were diagonal,
one can easily obtain the energy of incident electron through track length event by event.
However, these two plots show significant non-diagonal elements of the response matrices,
and hence it is impossible to obtain directly the incident energy through track length.
Especially, it is obvious that the track length in iron target is smeared widely
with energy above $30$ MeV, while for lead target the smearing begins with energy above $10$ MeV or so.
This is the consequence of electromagnetic shower.
Both simulations are exactly consistent with the expected critical energies of $\mathrm{Fe}$ and $\mathrm{Pb}$.
\end{multicols}

\ruleup
\begin{center}
\includegraphics[width=14 cm]{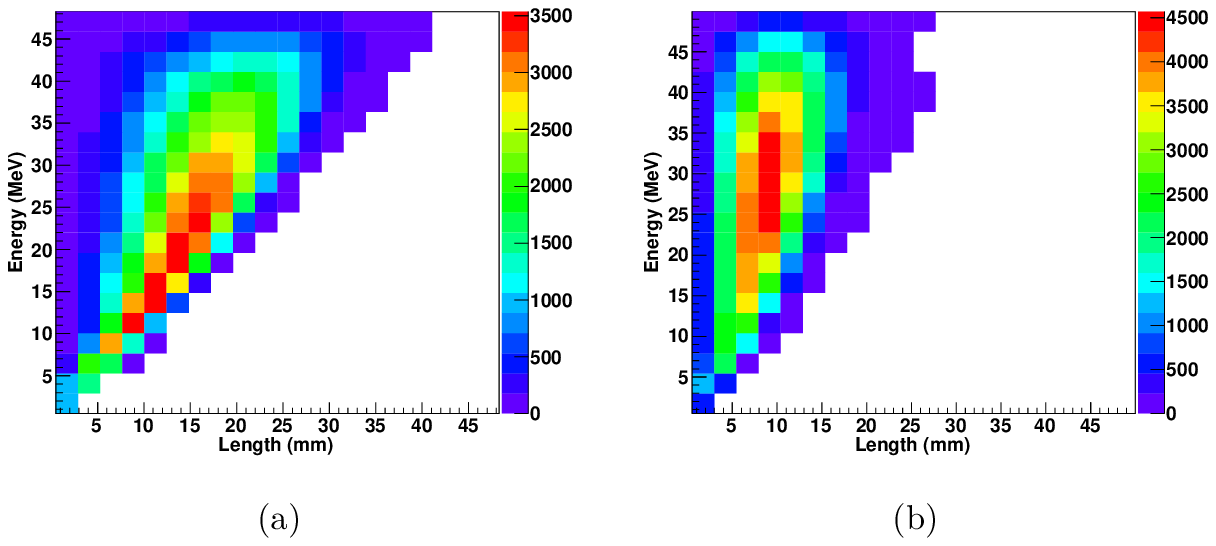}
 \figcaption{ \label{fig:responseM}
    Scattering plot of electron energy $E$ versus track length $L$
    for target of Fe (a) and Pb (b).
    Significant non-diagonal elements of response matrix exist due to electromagnetic shower above critical energy $E_c$.
    The energy spectrum of incident electrons is Michel distribution.
 }
\end{center}
\ruledown

\begin{multicols}{2}
To extract the physical part out of the unphysical oscillation,
the choice of regularization parameter is crucial,
which reflects a trade-off between bias and oscillation.
For the unfolding based on the SVD technique, the choice of regularization parameter
should be tuned for given distribution, number of bins and sample size.
As described in Ref.~\cite{SVD_Hocker},
the regularization parameter $kterm$ should be chosen by the plot of $\log|d|_i$ versus $i$,
where integer $i$ is from $1$ to the number of bins
and the $i$th component of the vector $d$ is the coefficient
in the decomposition of the measured (and re-scaled) histogram.
Only the first few terms, say $k$, of the decomposition should be significant
for a reasonably smooth measured distribution,
while the others correspond to contribution of quickly oscillating basis vectors,
and should be compatible with zero.
Therefore, one should see two separate patterns on the plot of $\log|d|_i$:
for small $i$, $|d_i|$ is significantly greater than $1$ and falls gradually
to a standard gaussian distribution for large $i$.
One should choose the regularization parameter $kterm$ to be the critical value $i=k$,
after which $d_i$'s are insignificant.

Fig.~\ref{fig:ddi}(a) shows the plot of $\log|d_i|$ versus $i$ with $5,000$ events accumulated
in $\mathrm{Fe}$ target, approximate data of one nominal year.
The parameter $kterm=5$ is chosen from this plot.
The unfolded energy spectrum of the $\nu_e$ induced electron
is shown in the upper plot of Fig.~\ref{fig:unfoldedE}(a),
while the lower plot shows the discrepancy between the unfolded spectrum and the true one,
where the latter is Michel distribution shaped by theoretical prediction of differential cross section.
With the unfolded energy spectrum of the induced electron,
one can obtain the spectrum of the $\nu_e$'s that interact with the $\mathrm{Fe}$ target using Eq.~(\ref{eq:energy_relation}).
Dividing this spectrum by Michel distribution gives
the differential cross section of $\nu_e$-nucleus scattering,
as shown in the upper plot of Fig.~\ref{fig:xsec}(a).
The lower plot in Fig.~\ref{fig:xsec}(a) shows the relative discrepancy
between the measured cross section and the true one.
One can see that the discrepancies and errors are huge for energy below $20$ MeV,
mainly because the cross section of low energy is too small and few events are collected in this range.
The discrepancy of the last bin is also large, around $30\%$.
For other bins, the discrepancies are controlled within $20\%$.
We checked this result by comparison with five other samples under the same condition.
It shows that for energy below $20$ MeV
the measured cross sections fluctuate significantly, therefore totally unreliable.
While the discrepancies between $20$ and $50$ MeV are well controlled within $20\%$.
Increasing the sample size to be $10000$ slightly improves the behavior between $20$ and $50$ MeV,
but helps less(if not nothing) to the region below $20$ MeV.
\end{multicols}
\ruleup
\begin{center}
\includegraphics[width=10 cm]{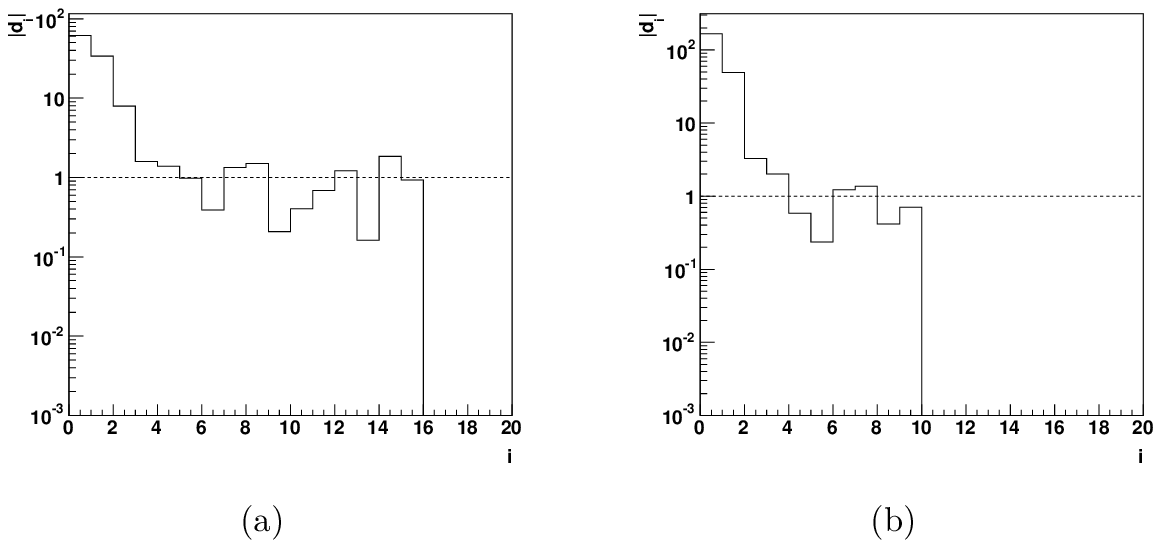}
 \figcaption{ \label{fig:ddi}
   Plots of $\log|d_i|$ versus $i$ for target of $\mathrm{Fe}$ (a) and $\mathrm{Pb}$ (b).
   The number of bins is $20$ for both targets, while the number of measured events is
   $5,000$ for $\mathrm{Fe}$ and $30,000$ for $\mathrm{Pb}$.
 }
\end{center}

\begin{center}
\includegraphics[width=10 cm]{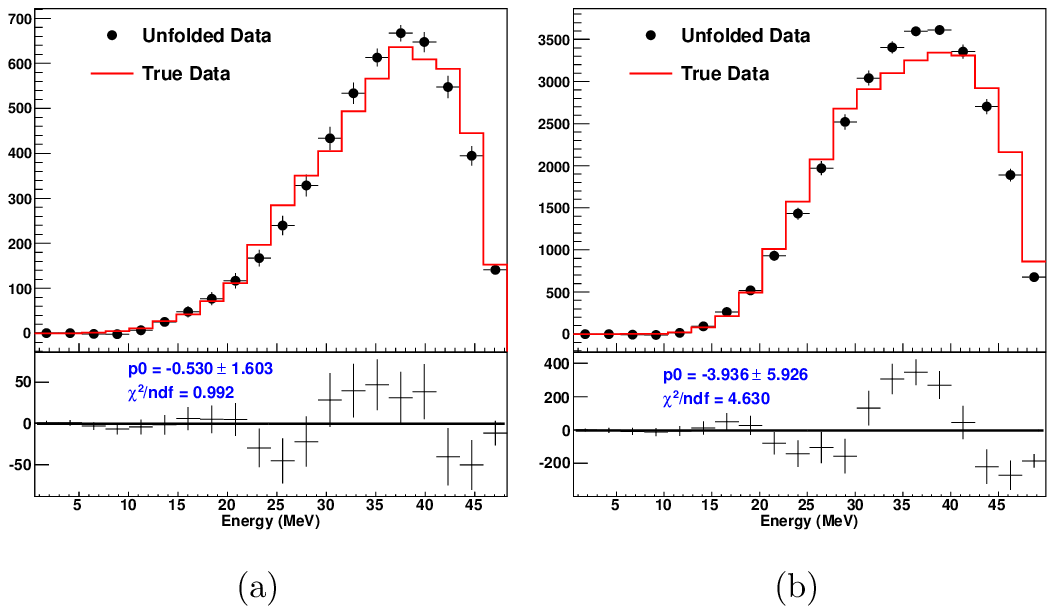}
 \figcaption{ \label{fig:unfoldedE}
   Comparison between the unfolded and true energy spectra of the $\nu_e$ induced electrons,
   The target is $\mathrm{Fe}$ for (a) and $\mathrm{Pb}$ for (b).
   The number of bins is $20$ for both targets, while the number of measured events is
   $5,000$ for $\mathrm{Fe}$ and $30,000$ for $\mathrm{Pb}$.
 }
\end{center}

\begin{center}
\includegraphics[width=10 cm]{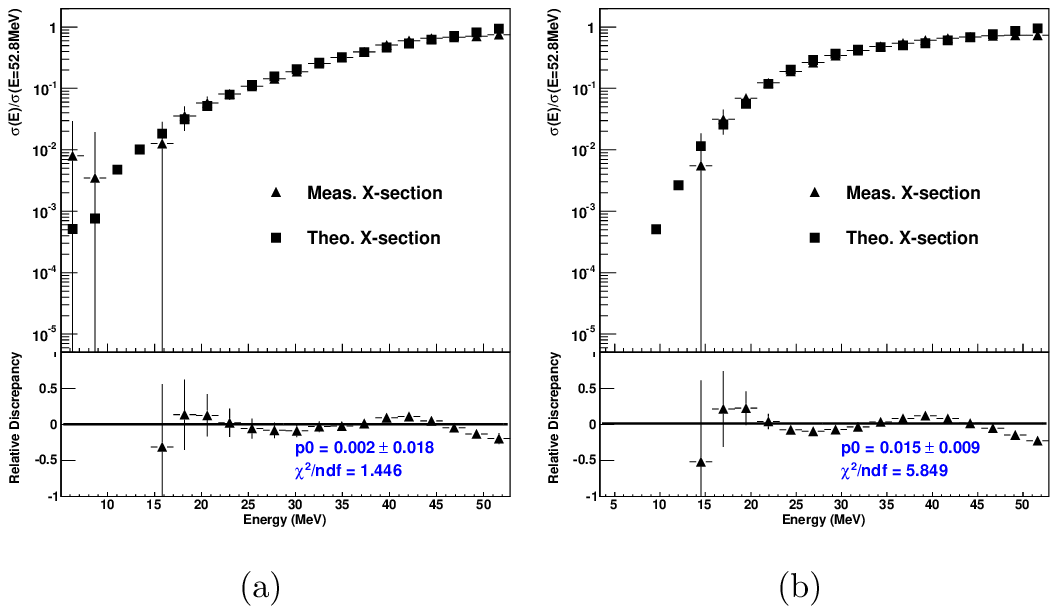}
 \figcaption{ \label{fig:xsec}
   Comparison between the measured and true differential cross sections
   of $\nu_e$-$\mathrm{Fe}$ (a) and $\nu_e$-$\mathrm{Pb}$ (b) scattering.
   The number of bins is $20$ for both targets, while the number of measured events is
   $5,000$ for $\mathrm{Fe}$ and $30,000$ for $\mathrm{Pb}$.
   $\sigma(E=52.8\mathrm{MeV})$ is $1.2\times10^{-39}\mathrm{cm}^2$ for $\mathrm{Fe}$
   and $1.3\times10^{-38}\mathrm{cm}^2$ for $\mathrm{Pb}$.
 }
\end{center}

\ruledown
\begin{multicols}{2}
For $\mathrm{Pb}$ target, the sample size is $30,000$.
Fig.~\ref{fig:ddi}(b) shows the plot of $\log|d_i|$ versus $i$, according to which,
the regularization parameter $kterm=4$ is chosen for $\mathrm{Pb}$.
The unfolded energy spectrum of the $\nu_e$ induced electron
is shown in the upper plot of Fig.~\ref{fig:unfoldedE}(b),
and the lower plot shows the discrepancy between the unfolded and true spectrum,
where the true one is Michel distribution shaped
by theoretically predicted cross section of $\nu_e$-$\mathrm{Pb}$ interaction.
The resulting differential cross section of $\nu_e$-$\mathrm{Pb}$ scattering
is shown in the upper plot of Fig.~\ref{fig:xsec}(b),
and the lower plot shows the relative discrepancy
between the measured differential cross section and the true one.
Similar to the case of $\mathrm{Fe}$ target,
one can see that the discrepancies and errors are large for energy below $22$ MeV
due to lack of events.
The discrepancy of the last bin is also large, around $30\%$.
For other bins, energy between $25$ and $50$ MeV,
the discrepancies are well controlled within $20\%$.
This result is compared with five other samples using the same regularization parameter.
It shows that for energy below $22$ MeV
the measured cross sections fluctuate significantly, therefore totally unreliable.
While the discrepancies between $22$ and $50$ MeV are well controlled below $20\%$.

\section{Conclusions}
With Monte Carlo simulation, we studied the achievable accuracy
for the $\nu_e$-nucleus scattering cross section measurement using an ideal segmented detector.
Two different target materials, $\mathrm{Fe}$ and $\mathrm{Pb}$, are investigated,
since they are most relevant to the evolution of supernova and experimental data of the cross section are still not available.
The energy spectrum of the $\nu_e$ induced electron can be well reconstructed
using the unfolding method, therefore makes it feasible
for the  measurement of differential cross section.
For $\mathrm{Fe}$ target, the cross sections between $20$ and $50$ MeV can be measured
within $20\%$ accuracy using $5,000$ events, the cross sections below $20$ MeV is however
totally unreliable since the cross section is too small in this region.
For $\mathrm{Pb}$ target, the cross sections between $22$ and $50$ MeV can be measured
within $20\%$ accuracy using $30,000$ events, the cross sections below $22$ MeV are
not reliable due to low statistics.

\acknowledgments{
 We are grateful for the fruitful discussions with Dr. X. Zhu and Dr. Y. Lin for helpful discussions.
This work is partly supported by the National Natural Science Foundation of China under contract number 10875062.
}
\end{multicols}

\begin{multicols}{2}

\end{multicols}


\clearpage

\end{document}